\documentclass[preprint]{elsarticle}

\usepackage{amssymb}
\usepackage{bm}
\usepackage{amsmath}
\usepackage{caption}
\usepackage{cleveref}
\usepackage{multirow}
\usepackage{color,soul}
\biboptions{compress}

\begin{document}
\captionsetup[figure]{labelfont={bf},name={Fig.},labelsep=period}
\captionsetup[table]{labelfont={bf},name={Table.},labelsep=period}
\sethlcolor{yellow}
\setstcolor{red}
\soulregister\cite7 % 针对\cite命令

\begin{frontmatter}

\title{A conservative level-set method based on a posterior mass correction preserving distance property for incompressible multiphase flows simulations}

\author{Tian Long}%\corref{cor1}\fnref{label3}}
%\cortext[cor1]{I am corresponding author}
%\fntext[label3]{I also want to inform about\ldots}
%\fntext[label4]{Small city}

%\ead{author.one@mail.com}
%\ead[url]{author-one-homepage.com}
\author{Jinsheng Cai}

\author{Shucheng Pan\corref{cor1}}
%address[label1]{Some University}
\ead{shucheng.pan@nwpu.edu.cn}

\cortext[cor1]{Corresponding author}

%\ead{author.three@mail.com}
\address{School of Aeronautics, Northwestern Polytechnical University, Xi'an, 710072, PR China}

\begin{abstract}
As one of the most popular interface-capturing methods, the level-set method is inherently non-conservative, and its evolution usually leads to unphysical mass gain/loss. In this paper, a novel conservative level set method is developed for incompressible multiphase flows simulations. A posterior mass correction is performed by introducing a small perturbation to the level-set field, which is solved via the Newton method. Unlike in previous researches, the signed distance property of the level-set function is exactly preserved after the present mass correction. Moreover, this method can be easily generalized from 2D to 3D. The influence for the computational efficiency is slight as the correction does not need to be applied at every time step. Various benchmark cases involving pure interface-evolution problems and multiphase flows problems are considered to validate the present method. For all cases, the accuracy and efficiency of the original method and the present method are quantitatively compared. It is observed that, with negligibly extra cost, the conservation error is reduced to the order of machine accuracy by the present method, which indicates its potential applications in complex multiphase flows simulations.
\end{abstract}


\end{frontmatter}

% \linenumbers

% main text
\section{Introduction}
\label{sec1}
%\subsection{Sample subsection}
%\label{subsec1}
Over the last decades, many numerical methods have been developed for the simulation of multiphase flows due to its widely practical applications, such as ink jet printing \cite{yu2005coupled}, bubble column reactors \cite{Olmos2001Numerical}, gas turbine engines \cite{rachner2002modelling}, etc. In general, these methods can be divided into two classes based on the treatment of the material interface separating different fluids, namely interface tracking methods and interface capturing methods. In the interface tracking methods, the interface is explicitly tracked by various markers, e.g., the mesh used in the arbitrary Lagrangian Eulerian (ALE) method \cite{Hirt1974arbitrary,Ling2010numerical}, which evolves and deforms with background flows, and the Lagrangian particles used in the front-tracking method \cite{unverdi1992front}. Through the explicit interface representations, the multiphase flows can be simulated with very high accuracy. Nevertheless, these methods are not suitable for complex problems as the typological changes have to be handled manually, which is challenging and computationally expensive. In contrast, the interface capturing methods, in which the interface is implicitly captured through an auxiliary function, can cope with the interface deformations and topological changes in an automatical way. One such method is the volume of fluid (VOF) method \cite{rudman1997volume,Rider1997Reconstructing,Gueyffier1999Volume,Scardovelli2000Analytical,Pilliod2004Second}, which describes the interface through the color function representing the fraction of the liquid volume in each computational cell. Although it is intrinsically mass-conserving, this method suffers from the inaccurate computation of geometrical variables such as normals and curvature, owing to the discontinuous feature of the color function. Another popular interface capturing method is the level-set method proposed by Osher and Sethian \cite{osher1988fronts}, which represents the interface with the zero contour of the so-called level-set function defined by the signed distance to the interface. The interface evolution and the related geometrical variables can be computed accurately since the level-set function is Lipschitz-continuous. In recent years, the level-set method is increasingly used in the simulation of multiphase flows \cite{gibou2018review} due to its simplicity and high efficiency. However, unphysical mass gain/loss will occur when it is applied to incompressible multiphase flows. The main reasons for the missing conservation property are twofold. First, the level-set filed will be evolved via solving an advection equation, whose discretization will unavoidably lead to numerical errors. Second, additional conservation errors will be introduced because of the artificial alteration of the interface in the reinitialization procedure \cite{sussman1994level}, which is needed to maintain the signed distance property of the level-set function.

To achieve mass conservation in the level-set method, many numerical approaches have been proposed, which can be classified into the following five categories. The first strategy is the most straightforward one, i.e., to reduce numerical errors via using high-order discretization schemes \cite{salih2009numericalstudies,xiu2001semilagrangian}, minimizing the displacement of the zero level-set during reinitialization \cite{nourgaliev2007highfidelity,russo2000remark}, or employing a local mesh refinement \cite{min2007secondadaptive,herrmann2008balanced,nourgaliev2005improving}. Nevertheless, these methods can only mitigate (but not eliminate) the conservation errors. The second approach is to modify the definition of the level-set function to reduce the overall mass loss. One such example is the hyperbolic-tangent level-set method\cite{olsson2005conservative,olsson2007conservative}, which, however, shows a tendency to generate unphysical pieces of fluids called flotsams/jetsams breaking off where the mesh is under-resolved \cite{desjardins2008accurate}. Thirdly, to improve the mass conservation, some methods have coupled the level-set method with other interface descriptions, e.g., the coupled level-set/volume of fluid method (CLSVOF) \cite{M2007Coupling,Sussman2000Coupled}, the hybrid particle level-set method (HPLS)\cite{Enright2002Hybrid,Wang2009improved}, and the level-set volume constraint method (HLSVC) \cite{wang2012hybrid}. Although these hybrid methods show a remarkable improvement on the mass conservation, the simplicity of the original level-set method is also lost. The fourth remedy is enforcing a volume constraint during the evolution of the interface. For instance, in the method of Yuan et al. \cite{yuan2018simple}, the volume constraint is added as a source term in the advection equation. However, they assumed a uniform distribution of the source term in the whole domain, which lacks physical validity. The last approach is the posterior mass correction, in which the absolute mass loss within the computational domain will be redistributed. Examples of such methods include the mass-conserving level-set method (MCLS) \cite{van2005massconserving,van2008computing}, the curvature-based mass-redistribution method \cite{luo2015massconserving} and the interface-correction level-set method (ICLS) \cite{ge2018efficient}. Since these methods cannot preserve the signed distance property of the level-set function, a reinitialization step is needed after the correction, in which new conservation errors will arise.  

In this paper, a novel conservative level-set method for incompressible multiphase flows simulations is proposed. Inspired by the stimulus-response algorithm of Han et al. \cite{han2015scale}, we have developed a novel posterior mass correction procedure which does not violate the signed distance property of the level-set function. As the correction step is not implemented at every time step, the extra computational cost is negligible. We also emphasize that the present method is strictly conservative, i.e., the conservation error is of the order of machine accuracy. The remainder of this paper is organized as follows. In Section 2, we briefly review the numerical scheme for incompressible multiphase flows with the traditional level-set method. Subsequently in Section 3 we introduce the novel conservative level-set method. A number of numerical tests are carried out to validate the present method in Section 4, followed by the concluding remarks in Section 5.
\section{Numerical method}
\label{sec2}
\subsection{Governing equations}
\label{subsec2.1}
The non-dimensional governing equations for incompressible multiphase flows are
\begin{subequations}
\begin{align}
&\nabla \cdot \bm{V} = 0, \label{eq:mass equation}\\
&\frac{\partial \bm{V}}{\partial t} + \bm{V}\cdot\nabla\bm{V} = \frac{1}{\rho} \left[ -\nabla p + \frac{1}{Re}\nabla\cdot\left(\mu(\nabla\bm{V}+\nabla\bm{V}^T)\right) \right]+\frac{1}{Fr}\bm{g}, \label{eq:momentum equation}
\end{align}
\label{eq:NSequations}
\end{subequations} 
where $\bm{V} = (u,v,w)$ is the velocity vector, $\rho$ the density, $p$ the pressure, $\mu$ the dynamic viscosity, and $\bm{g}$ the unit vector aligned with gravity. For viscous flows considered here, the velocity field and its tangential derivatives are continuous across the interface $\Gamma$ that separates two different fluids (denoted as fluid$1$ and fluid$2$). In contrast, the material properties are subject to a jump across the interface, i.e., $[\rho]_\Gamma = \rho_2 - \rho_1$ and $[\mu]_\Gamma = \mu_2 - \mu_1$, where $[.]_\Gamma$ represents a jump in the variable considered. The pressure jump caused by the viscosity discontinuity and surface tension reads
\begin{equation}
[p]_\Gamma = \frac{1}{We}\kappa + \frac{2}{Re}[\mu]_\Gamma \bm{n}^T\cdot\nabla\bm{V}\cdot\bm{n},
\label{eq:pressure jump}
\end{equation}
in which $\bm{n} = (n_x,n_y,n_z)$ is the interface noraml, and $\kappa$ is the curvature. The dimensionless parameters in Eqs. (\ref{eq:NSequations}) and (\ref{eq:pressure jump}), namely the Reynolds number $Re$, the Froude number $Fr$, and the Weber number $We$, are defined as
\begin{equation}
Re = \frac{\widetilde{\rho}\widetilde{U}\widetilde{L}}{\widetilde{\mu}}, \qquad Fr = \frac{\widetilde{U}^2}{\widetilde{g}\widetilde{L}}, \qquad We = \frac{\widetilde{\rho}\widetilde{U}^2\widetilde{L}}{\widetilde{\sigma}},
\label{eq:nondimensional parameters}
\end{equation}
where $\widetilde{\rho}$, $\widetilde{U}$, $\widetilde{L}$, $\widetilde{\mu}$, $\widetilde{g}$ and $\widetilde{\sigma}$ represent the reference density, velocity, length, dynamic viscosity, gravitational acceleration, and surface tension coefficient, respectively. Note that the surface tension coefficient $\widetilde{\sigma}$ is assumed to be constant over space and time in this paper.
\subsection{Interface representation}
\label{2.2}
In a two fluid system, the flow domain $\Omega$ is decomposed into two sub-domains $\Omega_1$ and $\Omega_2$ by an evolving interface $\Gamma(t)$, which is implicitly defined by the zero contour of the level-set function $\phi(\bm{x},t)$ representing the signed distance from the interface to $\bm{x}$. Without loss of generality, let $\phi > 0$ in $\Omega_1$ (fluid1) and $\phi < 0$ in $\Omega_2$ (fluid2). The interface movement is captured via solving the level-set equation \cite{fedkiw1999ghost},
\begin{equation}
\frac{\partial \phi}{\partial t} + \bm{V}\cdot\nabla\phi = 0,
\label{eq:level-set equation}
\end{equation}
where $\bm{V}$ is the background fluid velocity. With a given level-set function, the normal vector $\bm{n}$ and curvature $\kappa$ can be easily obtained by
\begin{equation}
\bm{n} = \nabla \phi, \quad \text{and} \quad \kappa = \nabla \cdot \frac{\nabla \phi}{|\nabla \phi|},
\label{eq:geometrical variables}
\end{equation}
respectively. In practice, a reinitialization procedure is required to enforce the signed distance property $|\nabla \phi = 1|$ which will be violated due to the deformations of the interface. Following Sussman et al. \cite{sussman1998improved}, the level-set function is reinitialized via iteratively solving a Hamilton-Jacobi equation,
\begin{equation}
\frac{\partial \phi}{\partial \tau} + S(\phi_0) (|\nabla \phi| - 1) = 0,
\label{eq:reinitialization}
\end{equation}
where $\tau$ is the pseudo time, and $S(\phi_0)$ is the sign function of the original level-set. Note that there is no need to implement the reinitialization at every time step \cite{luo2015massconserving}. Following Ref. \cite{ge2018efficient}, in the present study, we perform 10 iterations of Eq. (\ref{eq:reinitialization}) for every 20 time steps.
\subsection{Numerical scheme}
\label{2.3}
The solver we used employs a staggered uniform grid, in which the pressure $p$  and the level-set function $\phi$ are stored at cell centers while the velocity is stored at cell faces. For the solution of the Navier-Stokes equations, spatial distretization is performed by using the second-order accurate finite central difference scheme, and the projection method of Chorin \cite{chorin1968numerical} is used for time marching. 

Before updating the flow filed, $\phi^{n+1}$ is obtained through solving the level-set equation Eq. (\ref{eq:level-set equation}). The fifth-order accurate weighted essentially non-oscillatory (WENO5) scheme \cite{liu1994weighted} and the third-order accurate strong stability preserving (SSP) Runge-Kutta (RK3) scheme \cite{shu1988efficient} are used for the evaluation of $\nabla \phi$ and the temporal discretization, respectively. With $\phi^{n+1}$, the density and velocity fields can be updated by 
\begin{equation}
\begin{aligned}
&\rho^{n+1} = \rho_1 H_s(\phi^{n+1}) + \rho_2(1 - H_s(\phi^{n+1})), \\
&\mu^{n+1} = \mu_1 H_s(\phi^{n+1})  + \mu_2(1 - H_s(\phi^{n+1})),
\end{aligned}
\label{eq:smeared density and velocity}
\end{equation}
where 
\begin{equation}
H_s(\phi) = \begin{cases}
0   &\text{if}\ \phi <  \mathrm{-}\epsilon\\
\frac{1}{2} \left[ 1+\frac{\phi}{\epsilon}+\frac{1}{\pi}\sin(\frac{\pi\phi}{\epsilon}) \right] &\text{if}\ |\phi| \le \epsilon \\
1  &\text{if}\ \phi > \epsilon
\end{cases}
\label{eq:smeared out heaviside}
\end{equation}
is the smoothed Heaviside function with $\epsilon = 1.5 \Delta x$ being the half thickness of the smeared interface.

Let $\textbf{RU}^{n}$ denotes the right-hand side of the momentum equation Eq. (\ref{eq:momentum equation}) with the pressure gradient term excluded, which reads
\begin{equation}
\textbf{RU}^n = - \bm{V}^n\cdot\nabla\bm{V}^n + \frac{1}{\rho^{n+1}} \left[ \frac{1}{Re}\nabla\cdot\left(\mu^{n+1}(\nabla\bm{V}^{n}+(\nabla\bm{V}^n)^T)\right) \right] + \frac{1}{Fr}\bm{g}.
\label{eq:RHS for prediction}
\end{equation}
Then, by using the second-order accurate Adams-Bashforth scheme (AB2) for the temporal integration, the intermediate velocity $\bm{V}^*$ can be computed by
\begin{equation}
\bm{V}^* = \bm{V} + \Delta t \left( \frac{3}{2} \textbf{RU}^{n} - \frac{1}{2} \textbf{RU}^{n-1} \right).
\label{eq:intermediate velocity}
\end{equation}
Finally the divergence-free velocity $\bm{V}^{n+1}$ becomes
\begin{equation}
\bm{V}^{n+1} = \bm{V}^* - \frac{\Delta t}{\rho^{n+1}} \nabla p^{n+1},
\label{eq:real velocity}
\end{equation}
where the pressure filed $p^{n+1}$ is obtained by solving a Poisson equation,
\begin{equation}
\nabla \cdot \left( \frac{1}{\rho^{n+1}}\nabla p^{n+1} \right) = \frac{1}{\Delta t} \nabla \cdot \bm{V}^*.
\label{eq:poisson equation}
\end{equation}
When discretizing the pressure gradient term in Eq. (\ref{eq:poisson equation}), we treat the pressure jump Eq. (\ref{eq:pressure jump}) via the ghost fluid method (GFM) \cite{ge2018efficient}. Furthermore, the fast pressure-correction method \cite{dodd2014fast} is utilized to improve computational efficiency. 

\section{Conservative level-set method}
\label{sec3}
When applied to the simulations of multiphase flows, the classic level-set method inherently leads to the non-conservation of mass. The volume enclosed by the zero level-set will change over time, which implies the unphysical mass gain/loss in the context of incompressible flows. In this section, a novel conservative level-set method is developed to address this issue. We begin by analyzing the sources of volume gain/loss in the original level-set method. Let $V$, $V_1$ and $V_2$ represent the volume of $\Omega$, $\Omega_1$ and $\Omega_2$, respectively. The conservation of $V_2$ is chosen to describe the present method, which can be numerically computed by
\begin{equation}
V_{2} = \int_{\Omega} (1 - H_s(\phi)) d\Omega.
\label{eq:volume of omega2}
\end{equation}
Taking the total derivative of Eq. (\ref{eq:volume of omega2}) with respect to time $t$, we obtain
\begin{equation}
\frac{D V_2}{Dt} = \int_{\Omega} (1 - H_s(\phi)) \nabla\cdot\bm{V} d\Omega-\int_{\Omega} \delta_{s}(\phi)(\frac{\partial \phi}{\partial t} + \bm{V}\cdot\nabla\phi) d\Omega 
\label{eq:derivative of v2}
\end{equation}
where 
\begin{equation}
\delta_s(\phi) = H'_s(\phi)=\begin{cases}
\frac{1}{2\epsilon} \left[ 1 + \cos(\frac{\pi\phi}{\epsilon}) \right] &\text{if}\ |\phi| \le \epsilon \\
0.0  &\text{elsewise}
\end{cases}
\label{eq:smeared out Dirac delta}
\end{equation}
is the smoothed Dirac delta function. It can be observed from Eq. (\ref{eq:derivative of v2}) that the change rate of $V_2$ will be zero when the continuity equation Eq. (\ref{eq:mass equation}) and the level-set equation Eq. (\ref{eq:level-set equation}) are solved exactly. However, even with a high-order discretization scheme and a fine mesh, numerical errors are inevitably introduced, which results in the non-conservation $\delta V_2^{(d)}$. The additional conservation error $\delta V_2^{(r)}$ will arise from the reinitialization procedure as the the zero level-set can be altered due to numerical artifacts. By employing the truly upwind discretization \cite{russo2000remark} or the subcell reconstruction \cite{min2010reinitializing} near the interface, $\delta V_2^{(r)}$ can be mitigated to some extent, but not eliminated. For the previous conservative level-set methods employing the posterior mass correction, $\delta V_2^{(r)}$ is difficult to handle since they cannot preserve the signed distance property of $\phi$. Even though the volume is conserved after being remedied, new conservation error will be introduced in the consequent reinitialization process, see Refs. \cite{luo2015massconserving, ge2018efficient}. 

In the scale separation algorithm of Han et al. \cite{han2015scale}, a uniform positive or negative shift referred to as stimulus is applied for the level-set function to determine and separate non-resolvable interface structures. Inspired by this idea, a small perturbation $\varepsilon$ is used in this study to eliminate the total conservation error $\delta V = \delta V_2^{(d)} +  \delta V_2^{(r)}$. We also emphasize that, as $|\nabla (\phi+\varepsilon) |= |\nabla \phi|$ holds, this treatment does not violate the signed distance property. Let $V_2^0$ be the initial volume of $\Omega_2$, thus the target equation reads
\begin{equation}
\int_\Omega (1 - H_s(\phi + \varepsilon)) d \Omega = V_2 + \delta V = V_2^0.
\label{eq:target equation}
\end{equation}
This equation can be solved by the Newton's method. First, we define
\begin{equation}
f(\varepsilon) = \int_\Omega (1 - H_s(\phi + \varepsilon)) d \Omega - V_2^0 = 0,
\label{eq:target function}
\end{equation}
and the fomulars for the iteration procedure are given by
\begin{subequations}
\begin{align}
\varepsilon_{k+1} &=\varepsilon_{k} - \frac{f(\varepsilon_k)}{f'(\varepsilon_k)}, \label{eq:newton 1}\\
\varepsilon_0 &= \frac{\int_\Omega (1 - H_s(\phi ))d \Omega - V_2^0 }{ \int_\Omega  \delta_s(\phi) d \Omega} , \label{eq:newton 2} \\
f'(\varepsilon) &= -\int_\Omega  \delta_s(\phi + \varepsilon) d \Omega, \label{eq:newton 3}
\end{align}
\label{eq:newton iteration}
\end{subequations}
where $k$ indexes the iteration step. The relative volume error
\begin{equation}
E_r = \left|\frac{V_2 - V_2^0}{V_2^0}\right|
\label{eq:relative volume error}
\end{equation}
is used to control the iteration process and the termination happens when $E_r \leq E_t$. In this paper, the threshold $E_t$ is chosen to $1.0\times10^{-12}$, and 3-4 iterations are required for all cases we have computed. When considering boundary terms, e.g., fluid2 is the liquid phase flowing into the domain across an inlet boundary, we can perform the correction procedure in the same way by changing the right side of Eq. \eqref{eq:target equation} to $V_2^0 + U_{i} A_{i} t^{n}$, where $U_{i}$, $A_{i}$, and $t^{n}$ are the inflow velocity, the area of the inlet, and the current simulation time, respectively.

To elucidate the present method from a geometrical point of view, Eq. (\ref{eq:newton 2}) is reformulated as
\begin{equation}
\varepsilon_0 \int_\Omega  \delta_s(\phi) d \Omega = - \delta V.
\label{eq:geometrical explanation}
\end{equation}
When $\varepsilon_0$ is small, we have
\begin{equation}
H(\phi+\varepsilon_0) - H(\phi) \approx \delta_s(\phi) \varepsilon_0,
\label{eq:linear approximation}
\end{equation}
which indicates that the left side of Eq. (\ref{eq:geometrical explanation}) is the approximation of the volume enclosed by the iso-surfaces $\phi = 0$ and $\phi = -\varepsilon_0$. The iteration process can be considered as an adjustment to reimburse the approximate error. To improve numerical accuracy, in the present study, the volume integrals in Eqs. (\ref{eq:target equation}) - (\ref{eq:relative volume error}) are normalized by $V$. Moreover, implementing the correction procedure at every time step is superfluous. Additional computational efficiency can be achieved by choosing an appropriate correction frequency, which is discussed in Section. \ref{sec4.1.1}.
\section{Numerical results}
\label{sec4}
In this section, the proposed conservative level-set method is validated by both the pure interface-evolution problems and the multiphase flows problems. For all cases, the accuracy and efficiency of the present method are quantitatively compared with that of the original level-set method.
\subsection{Pure interface-evolution problems}
\label{sec4.1}
\subsubsection{Linear advection}
\label{sec4.1.1}
\begin{figure}[htbp]
\centering
\includegraphics[width=1.0\textwidth]{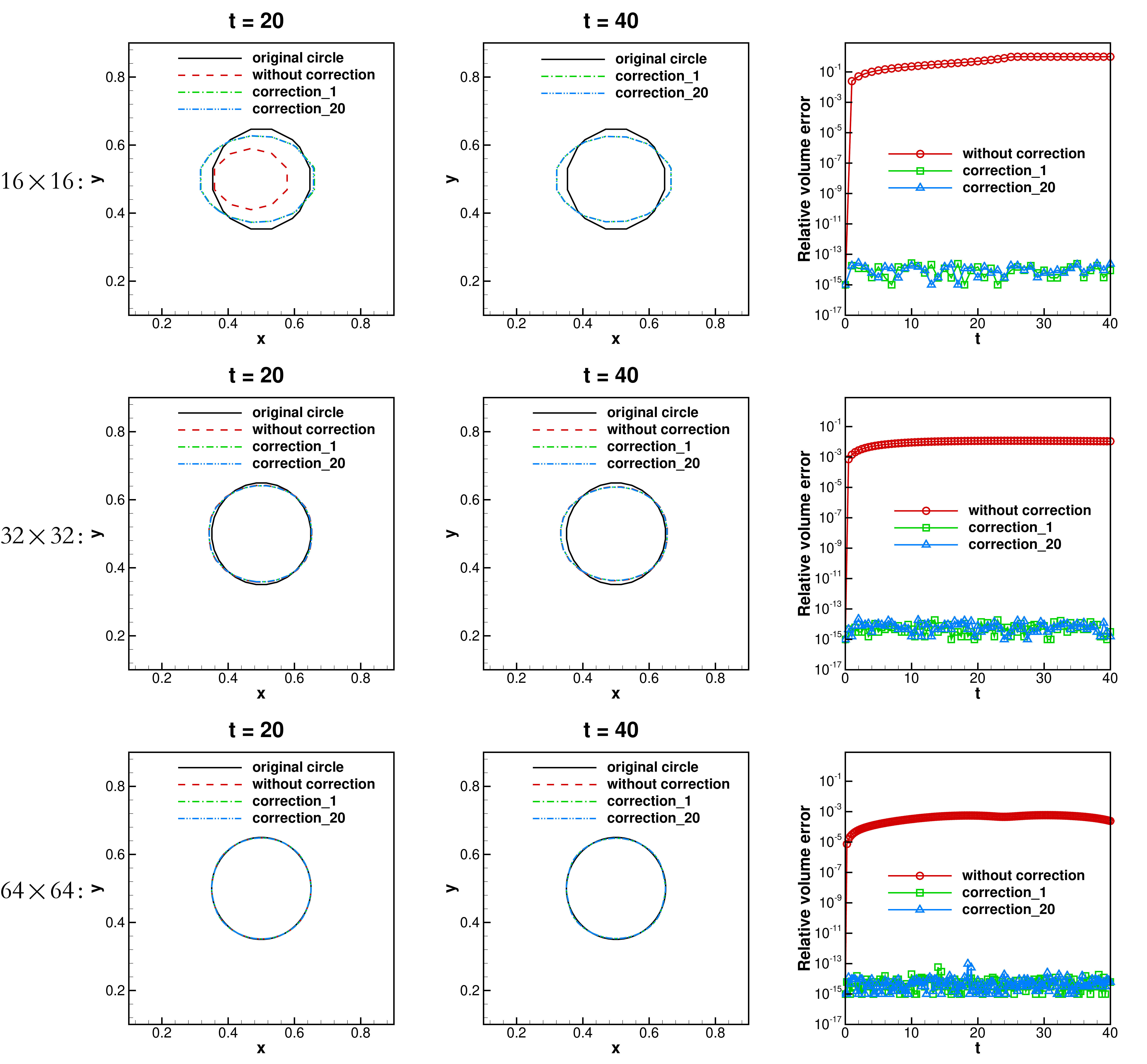}
\caption{Linear advection:  The interfaces at different time instants (the 1st and 2nd columns) and the relative volume errors (the 3rd column) for different resolutions.}
\label{Fig:linear advection}
\end{figure}

Firstly, as in \cite{van2005massconserving}, the linear advection case is considered to demonstrate the conservation. Driven by a constant velocity $u = 1$, a circle of radius $R = 0.1$ is initially placed at the center of a $[0,1] \times [0,1]$ square computational domain with periodic boundary conditions. This domain is discritized by a Cartesian grid, with resolutions increasing from $16 \times 16 $ to $64 \times 64$. This circle is advected for 40 periods with the time step $\Delta t$ of $0.8\Delta x$. To investigate the influence of the correction frequency, the correction procedure is applied for every time step and every 20 time steps. As shown in Table. \ref{Table:linear advection cpu costs}, the extra computational cost induced by the present method can be neglectable. On the coarsest mesh ($16\times16$), the interface indeed disappears due to the numerical dissipation of the original method, while it is preserved well with the present method, see Fig. \ref{Fig:linear advection}. In addition, the correction frequency does not affect the results. Thus, in the remaining cases, the correction procedure will be performed every 20 time steps for the sake of efficiency. Moreover, for different resolutions, the relative volume error is plotted versus the simulation time in a semi-logarithmic scale. It can be observed that the conservation error is significantly reduced with the present method. Note that the relative volume error obtained will be exactly zero at several time instants, which is set to $10^{-15}$ to be displayed in the logarithmic scale. 
%
%%%%%%%%%%%%%%%%%%%%%%%%%%%%%%%%%%%%%%%%%%%
\begin{table}
\centering
\caption{Linear advection: the CPU costs of the original method and the present method for different resolutions. For all cases, the simulations are performed on a server with an AMD EPYC 7742 64-core Processor (2.3 GHz).}
\resizebox{0.9\textwidth}{!}{
\begin{tabular}{cccc}
\hline
\multirow{2}*{Resolution}&\multicolumn{3}{c}{CPU costs ($ \rm{s} $)}\\
\cline{2-4}
&without correction&with correction (every 20 time steps)&with correction (every time step)\\
\hline
$16\times16$  (1 core)&0.1560&0.2043&0.2424\\
$32\times32$ (4 cores)&1.0682&1.1136&1.3068\\
$64\times64$ (4 cores)&6.1711&6.2657&6.6411\\
\hline
\end{tabular}
}
\label{Table:linear advection cpu costs}
\end{table}
%%%%%%%%%%%%%%%%%%%%%%%%%%%%%%%%%%%%%%%%%%%
%
%
%%%%%%%%%%%%%%%%%%%%%%%%%%%%%%%%%%%%%%%%%%%%%%
%%%%%%%%%%%%%%%%%%%%%%%%%%%%%%%%%%%%%%%%%%%%%%
%
\subsubsection{Zalesak's disk - rigid body rotation of a slotted disk}
\label{sec4.1.2}
\begin{figure}[htbp]
\centering
\includegraphics[width=1.0\textwidth]{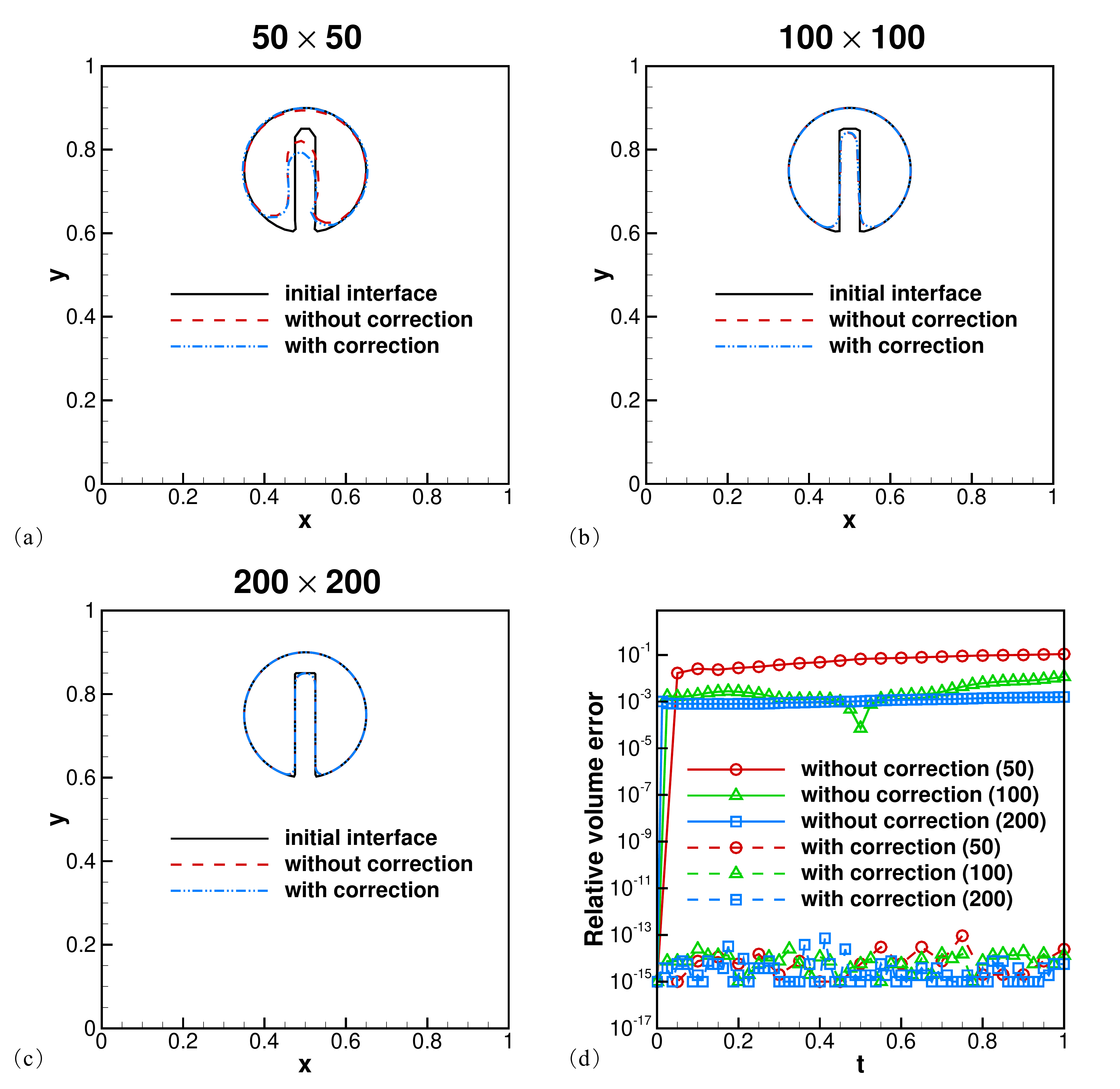}
\caption{Zalesak's disk: (a)-(c) the interfaces after one revolution for different resolutions, and (d) the relative volume errors for different resolutions.}
\label{Fig:Zalesak's disk}
\end{figure}

The second benchmark case is the rigid body rotation of the Zalesak's disk \cite{zalesak1979fully}, which has been extensively utilized to validate the interface-advection algorithms \cite{luo2015massconserving,ge2018efficient,chiu2019coupled}. Considering a unit square computational domain, with a radius of $0.15$, a slot length of $0.25$ and a slot width of $0.05$, the slotted circle centered at $(0.5, 0.75)$ is advected in a rotating velocity field given by
\begin{equation}
\begin{aligned}
u &= -2\pi(y - 0.5) \\
v &= 2\pi(x - 0.5).
\end{aligned}
\end{equation}
With $\Delta t = \Delta x/8$, the simulations are carried out on the grid with resolutions increasing from $50\times50$ to $200\times200$, see Table. \ref{Table:Zalesak's disk cpu costs}. The termination time is $t = 1$, after which the disk has recovered to its original position. In Fig. \ref{Fig:Zalesak's disk}(a)-(c), the final interface profiles obtained by the original method and the present method for different resolutions are compared, which indicate that the results of the present method are qualitatively better. As shown in Fig. \ref{Fig:Zalesak's disk}(d), with the mass correction procedure, the relative volume errors exhibit an order of machine accuracy for all resolutions.  
%
%%%%%%%%%%%%%%%%%%%%%%%%%%%%%%%%%%%%%%%%%%%%%%
%%%%%%%%%%%%%%%%%%%%%%%%%%%%%%%%%%%%%%%%%%%%%%
%
%
%%%%%%%%%%%%%%%%%%%%%%%%%%%%%%%%%%%%%%%%%%%
\begin{table}
\centering
\caption{Zalesak's disk: the CPU costs of the original method and the present method for different resolutions.}
\resizebox{0.6\textwidth}{!}{
\begin{tabular}{ccc}
\hline
\multirow{2}*{Resolution}&\multicolumn{2}{c}{CPU costs ($ \rm{s} $)}\\
\cline{2-3}
&without correction&with correction\\
\hline
$50\times50$  (4 cores)&0.6640&0.6818\\
$100\times100$ (16 cores)&4.3795&4.4153\\
$200\times200$ (16 cores)&17.271&17.407\\
\hline
\end{tabular}
}
\label{Table:Zalesak's disk cpu costs}
\end{table}
%%%%%%%%%%%%%%%%%%%%%%%%%%%%%%%%%%%%%%%%%%%
%
\subsubsection{Deformed interface in a vortex flow}
\label{sec4.1.3}
\begin{figure}[htbp]
\centering
\includegraphics[width=1.0\textwidth]{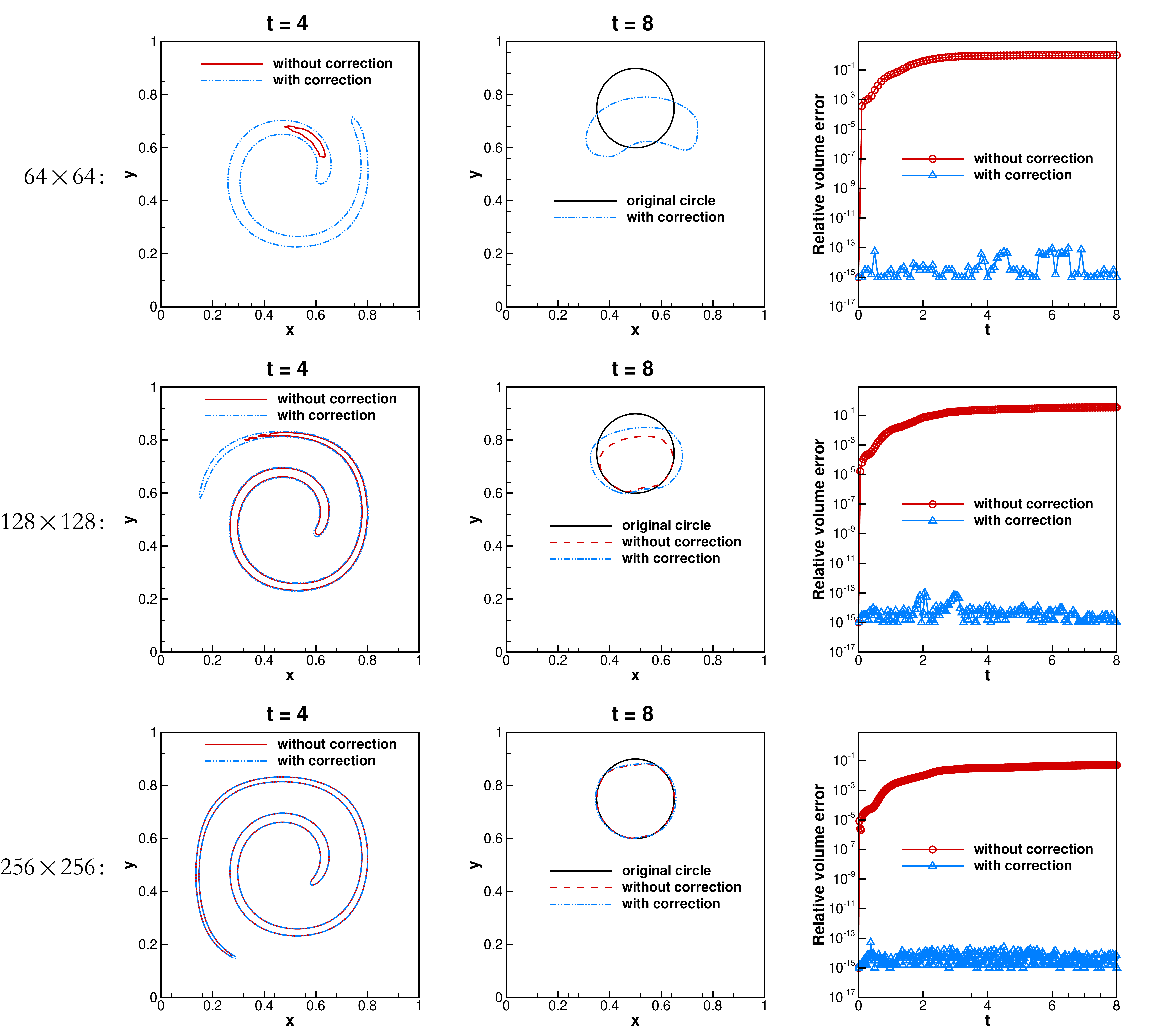}
\caption{Deformed interface in a vortex flow (2D): The interfaces at different time instants (the 1st and 2nd columns) and the relative volume errors (the 3rd column) for different resolutions.}
\label{Fig:vortex deformation 2D}
\end{figure}

Proposed by Bell et al. \cite{bell1989second}, the 2D reversed single vortex case is considered to assess the ability of the present method for capturing thin filaments. The computational domain is a unit square with a circle of radius $R = 0.15$, which is initially located at $(0.5,0.75)$ and deforms in a vortex flow. As in Refs. \cite{luo2015massconserving, ge2018efficient}, the velocity field is given by
\begin{equation}
\begin{aligned}
u&=-2 \sin (\pi x)^{2} \sin (\pi y) \cos (\pi y) \cos (\pi t/T), \\
v&=2 \sin (\pi y)^{2} \sin (\pi x) \cos (\pi x) \cos (\pi t/T),
\end{aligned}
\end{equation}
where $T = 8$ is the period of a reversing vortex flow. Deforming with the background velocity, the circle will be stretched into a thin filament at first. The largest deformation happens at $t = T/2$, and then the filament is pulled in reverse so that the initial circle is recovered at $t = T$. With $\Delta t = 0.32 \Delta x$, the simulations are conducted on the uniform grid with resolutions increasing from $64\times64$ to $256\times256$, see Table. \ref{Table:vortex deformation cpu costs}. The interfaces at $t = T/2$ and $t = T$ obtained by the original method and the present method are compared in Fig. \ref{Fig:vortex deformation 2D}. On the coarsest mesh of $64 \times 64$, the interface obtained with the original method is heavily dissipated, which totally vanishes at $t = T$. In contrast, with the same resolution, satisfying results are obtained by employing the present method. For all resolutions, the thin filament is better captured by the present method. The conservation is confirmed by the time history of the relative volume error, see the third column in Fig. \ref{Fig:vortex deformation 2D}.

\begin{figure}[htbp]
\centering
\includegraphics[width=1.0\textwidth]{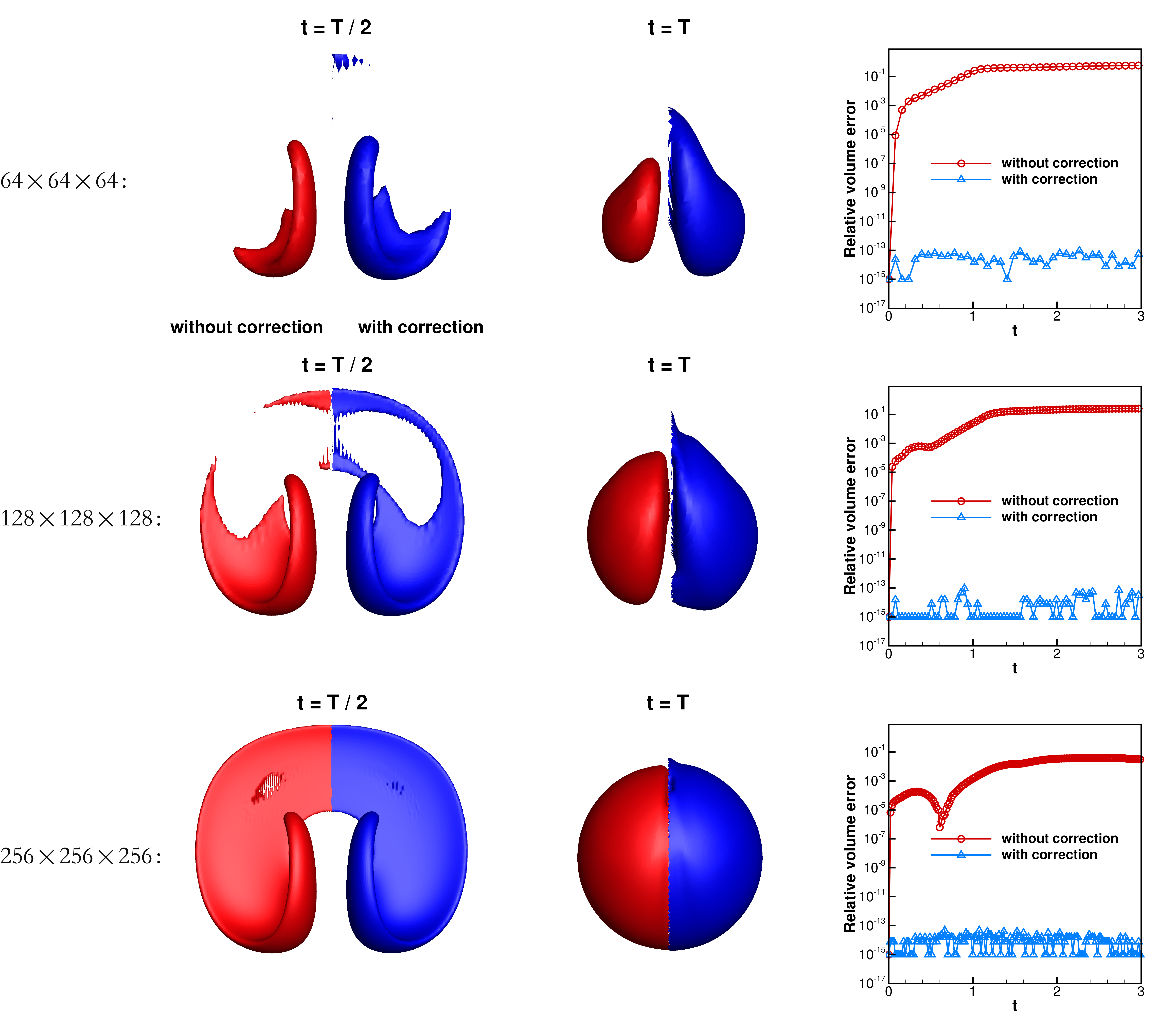}
\caption{Deformed interface in a vortex flow (3D): The interfaces at different time instants (the 1st and 2nd columns) and the relative volume errors (the 3rd column) for different resolutions.}
\label{Fig:vortex deformation 3D}
\end{figure}

The 3D version of this case is considered on the grid with increasing resolutions. Following LeVeque et al. \cite{leveque1996high}, a sphere of radius $R = 0.15$ is placed at $(0.35,0.35,0.35)$ in a unit cubic domain with the velocity field given by
\begin{equation}
\begin{aligned}
u&=2 \sin ^{2}(\pi x) \sin (2 \pi y) \sin (2 \pi z) \cos (\pi t / T), \\
v&=-\sin ^{2}(\pi y) \sin (2 \pi x) \sin (2 \pi z) \cos (\pi t / T), \\
w&=-\sin ^{2}(\pi z) \sin (2 \pi x) \sin (2 \pi y) \cos (\pi t / T),
\end{aligned}
\end{equation}
where $T$ is set to $3$. The time step is chosen as $\Delta t = \Delta x / 4$. It can be observed from Table. \ref{Table:vortex deformation cpu costs} that the CPU cost of the present method is almost the same as that of the original method. Note that the conservation error is greatly reduced with the present method, see Fig. \ref{Fig:vortex deformation 3D}.

%
%%%%%%%%%%%%%%%%%%%%%%%%%%%%%%%%%%%%%%%%%%%
\begin{table}
\centering
\caption{Deformed interface in a vortex flow: the CPU costs of the original method and the present method for different resolutions.}
\resizebox{0.8\textwidth}{!}{
\begin{tabular}{cccc}
\hline
&\multirow{2}*{Resolution}&\multicolumn{2}{c}{CPU costs ($ \rm{s} $)}\\
\cline{3-4}
&&without correction&with correction\\
\hline
\multirow{3}*{2D}
&$64\times64$  (4 cores)&3.0355&3.8993\\
&$128\times128$ (4 cores)&23.822&24.786\\
&$256\times256$ (16 cores)&96.666&99.107\\
\hline
\multirow{3}*{3D}
&$64\times64\times64$  (4 cores)&15.234&15.787\\
&$128\times128\times128$ (32 cores)&56.776&57.918\\
&$256\times256\times256$ (64 cores)&767.64&775.62\\
\hline
\end{tabular}
}
\label{Table:vortex deformation cpu costs}
\end{table}
%%%%%%%%%%%%%%%%%%%%%%%%%%%%%%%%%%%%%%%%%%%
%
%
%%%%%%%%%%%%%%%%%%%%%%%%%%%%%%%%%%%%%%%%%%%%%%
%%%%%%%%%%%%%%%%%%%%%%%%%%%%%%%%%%%%%%%%%%%%%%
%

%
%%%%%%%%%%%%%%%%%%%%%%%%%%%%%%%%%%%%%%%%%%%%%%
%%%%%%%%%%%%%%%%%%%%%%%%%%%%%%%%%%%%%%%%%%%%%%
%
\subsection{Multiphase flows problems}
\label{sec4.2}
\subsubsection{Falling droplet}
\label{sec4.2.1}
\begin{figure}[htbp]
\centering
\includegraphics[width=1.0\textwidth]{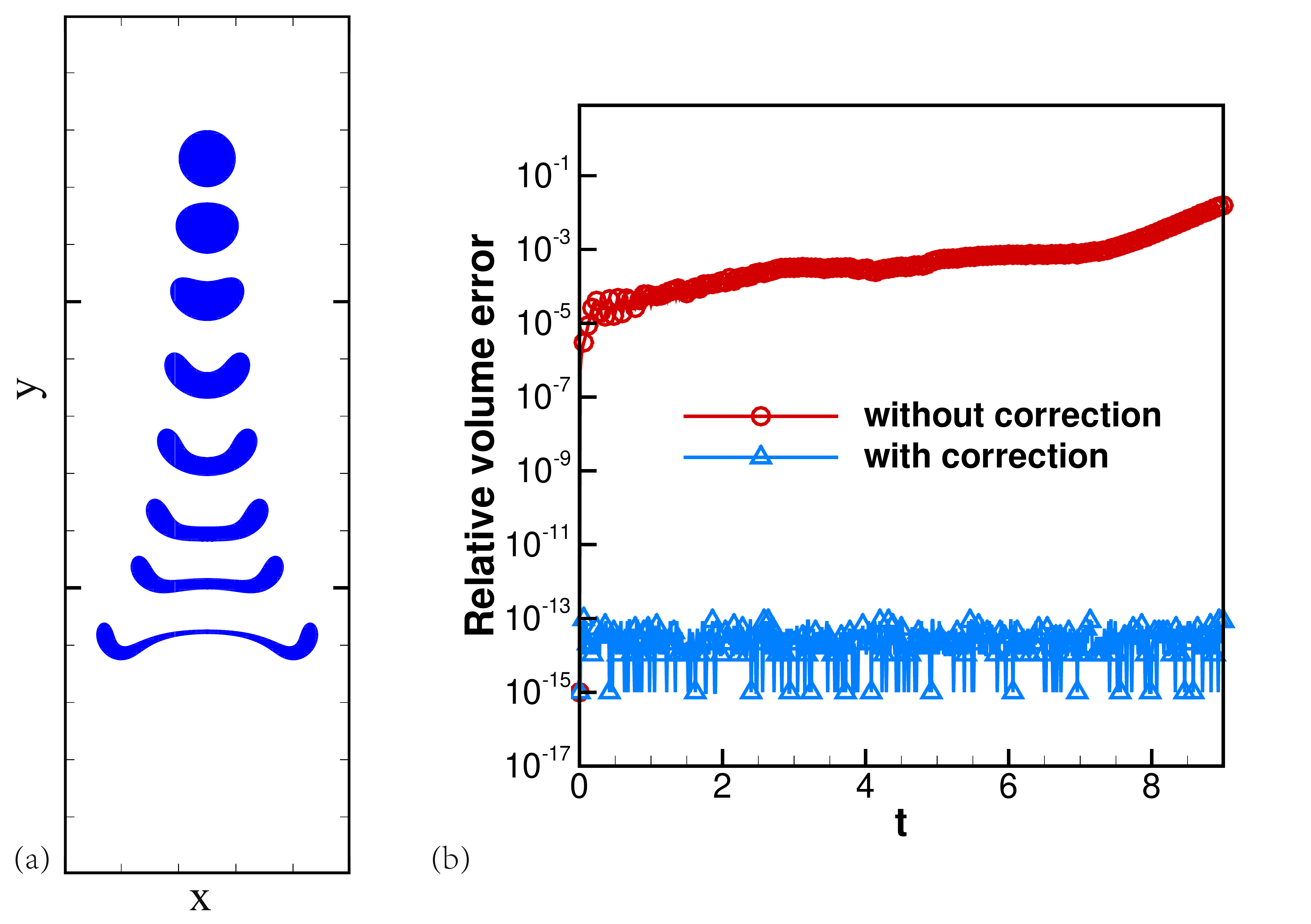}
\caption{Falling droplet: (a) the evolving interfaces of the falling droplet obtained by the present method at different time instants, which are corresponding to $t = 0, 2, 3, 4, 5, 6, 7, 9$ from top to bottom, respectively, and (b) the relative volume error over time.}
\label{Fig:drop_falling}
\end{figure}

We consider a droplet falling in air under gravity effect \cite{fakhari2016mass,long2021accelerated} here to validate the present method for multiphase flows problems. Within a rectangular domain of $[0,5]\times[0,15]$, a circular liquid drop with a diameter of $D = 1$ is initially placed at $(2.5,12.5)$ and then falls due to the gravity in the y-direction (see the coordinate axis in Fig. \ref{Fig:drop_falling}(a)). This case can be characterized by the density ratio $\bar{\rho} = \widetilde{\rho}_l/\widetilde{\rho}_a$, the Bond number $Bo = \widetilde{g} (\widetilde{\rho}_l - \widetilde{\rho}_a)\widetilde{D}^2/\widetilde{\sigma}$, and the gravity Reynolds numbers of liquid and air,
\begin{equation}
\begin{aligned}
&Re_l = \frac{\sqrt{\widetilde{g} \widetilde{\rho}_l(\widetilde{\rho}_l-\widetilde{\rho}_a)\widetilde{D}^3}}{\widetilde{\mu}_l}, \\
&Re_a = \frac{\sqrt{\widetilde{g} \widetilde{\rho}_a(\widetilde{\rho}_l-\widetilde{\rho}_a)\widetilde{D}^3}}{\widetilde{\mu}_a}.
\end{aligned}
\end{equation}
Following Refs. \cite{fakhari2016mass,long2021accelerated}, we set these characteristic parameters as $\bar{\rho} = 10$, $Bo = 100$, $Re_l = 20$ and $Re_a = 10$, which yields the following initial conditions,
\begin{equation}
\left\{
\begin{aligned}
\rho&=1,\ u = 0, \ v = 0,\  p = 1,\ \mu = 1 \quad \text{air},\\
\rho&=10,\ u = 0, \ v = 0,\ p = 1,\ \mu = 2\sqrt{10} \quad \text{liquid\ drop},\\
\phi&=-0.5+\sqrt{(x-2.5)^{2}+(y-12.5)^{2}}\quad\text{ level\ set},\\
Re &=6.635,\ We = 10,\ Fr = 1 \quad\text{ dimensionless parameters}.   \\ 
\end{aligned}
\right.
\end{equation}
For this case, the termination time is $t = 9.0$, with the time step being $\Delta t = 5.0\times10^{-4}$. The computational domain is discretized on a grid involving $120\times360$ cells. The no-slip wall boundary condition is employed for the top and bottom sides while the free-slip wall boundary condition is applied for the left and right sides. As shown in Fig. \ref{Fig:drop_falling}(a), expanding along the x-direction, the droplet gradually deforms into a bag shape similar with that in Fig. 9 of Ref. \cite{fakhari2016mass} and Fig. 21 of Ref. \cite{long2021accelerated}, and undergoes the bag breakup \cite{han1999secondary,jalaal2012fragmentation} at $t = 9$. By using eight cores, the CPU costs of the original method and the present method are $194.256\ \rm{s}$ and $196.332\ \rm{s}$, which indicate that the present method is as efficient as the original method. Note that the relative volume error is reduced to the order of machine accuracy via employing the present method, see Fig. \ref{Fig:drop_falling}(b). 
%
%%%%%%%%%%%%%%%%%%%%%%%%%%%%%%%%%%%%%%%%%%%%%%
%%%%%%%%%%%%%%%%%%%%%%%%%%%%%%%%%%%%%%%%%%%%%%
%
\subsubsection{Rayleigh-Taylor instability}
\label{sec4.2.2}
\begin{figure}[htbp]
\centering
\includegraphics[width=1.0\textwidth]{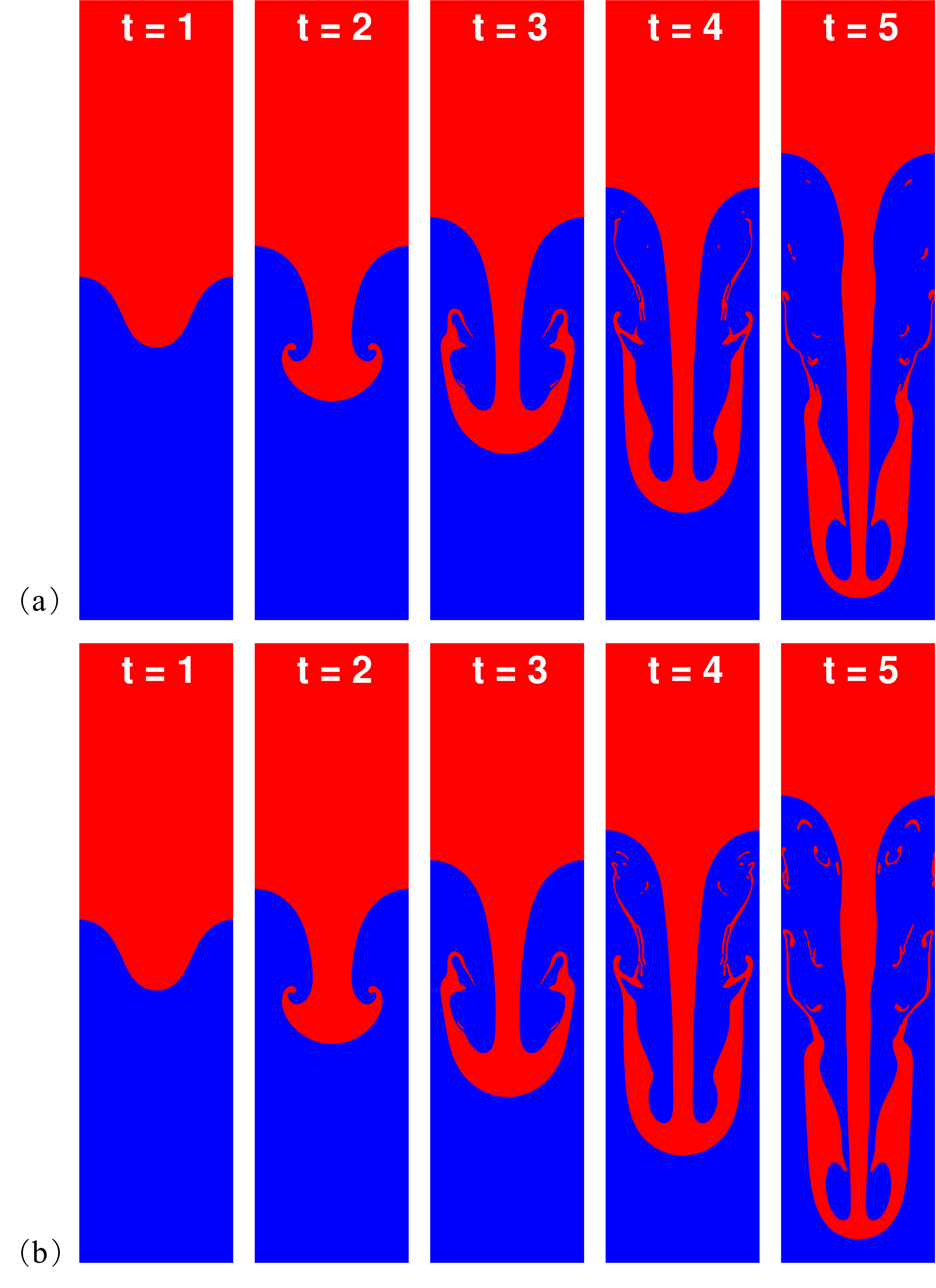}
\caption{Rayleigh-Taylor instability: the evolution of the interface obtained by (a) the original method and (b) the present method.}
\label{Fig:rt_instability interface}
\end{figure}

\begin{figure}[htbp]
\centering
\includegraphics[width=1.0\textwidth]{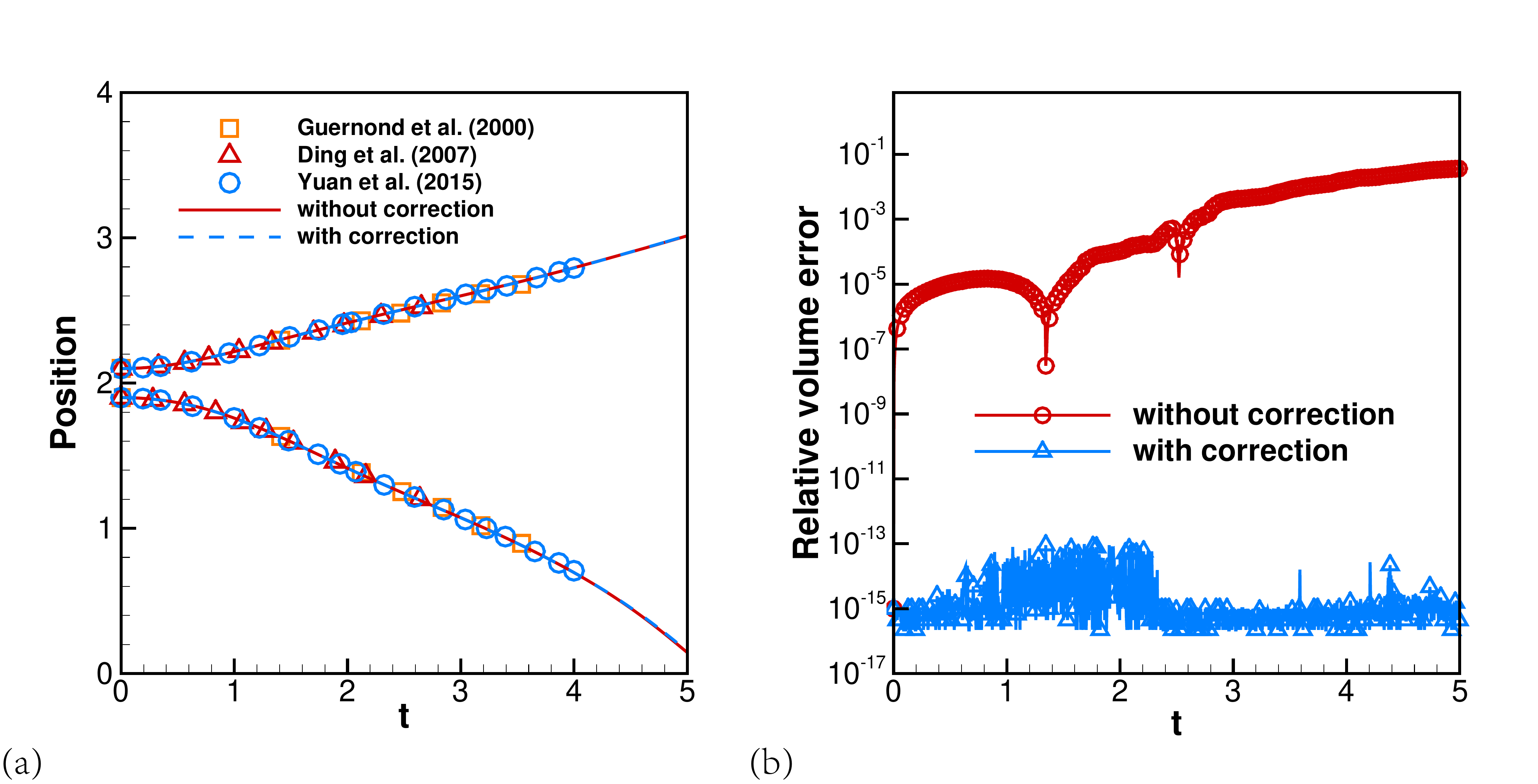}
\caption{Rayleigh-Taylor instability: (a) the y-coordinates of the tips of the falling and rising fluid, and (b) the relative volume error over time.}
\label{Fig:rt_instability position and error}
\end{figure}

Involving large deformations of the interface, the Rayleigh-Taylor instability problem with high Reynolds number is considered to assess the present method. In the absence of surface tension, a heavy fluid is located upon a light fluid with a small cosine perturbation along the interface, in which the Rayleigh-Taylor instability is induced. The density difference between two fluids is characterized by the Atwood ratio
\begin{equation}
At = \frac{\widetilde{\rho}_h - \widetilde{\rho}_l}{\widetilde{\rho}_h+\widetilde{\rho}_l},
\end{equation}
which is set to $0.5$ according to Refs. \cite{yuan2018simple,guermond2000projection,ding2007diffuse}. The flow field is solved in a rectangular domain of $[0,2]\times[0,4]$ with the initial conditions given by
\begin{equation}
\left\{
\begin{aligned}
\rho&=1,\ u = 0, \ v = 0,\  p = 1,\ \mu = 1 \quad \text{heavy\ fluid},\\
\rho&=1/3,\ u = 0, \ v = 0,\ p = 1,\ \mu = 1 \quad \text{light\ fluid},\\
\phi&=y - 2.0 - 0.1 \cos(2\pi y)\quad\text{ level\ set},\\
Re &=3000,\ Fr = 1 \quad\text{ dimensionless parameters}.   \\ 
\end{aligned}
\right.
\end{equation}
The time step and the termination time are set to $\Delta t = 2.0\times10^{-4}$ and $t = 5$, respectively. With the no-slip wall boundary condition imposed in the x-direction and the periodic boundary condition enforced in the y-direction, the simulations are carried out on a $200\times800$ grid using eight cores.  The computational time of the original method and the present method are $1279.68\ \rm{s}$ and  $1281.87\ \rm{s}$, respectively, which confirm the high efficiency of the present method. As shown in Fig. \ref{Fig:rt_instability interface}, the heavy fluid is falling down due to gravity with two counter-rotating vorticies rolled up along the interface. Growing with time, the two vortices are shed and a pair of secondary vortices are formed at the tails of the roll-ups. It can be observed that more details are captured with the present method. Fig. \ref{Fig:rt_instability position and error}(a) depicts the time history of the y-coordinates for the top of the rising fluid and the bottom of the falling fluid. Quantitatively, the numerical results of the present method are in good agreement with the previous simulations \cite{yuan2018simple,guermond2000projection,ding2007diffuse}. In addition, with the present method, the relative volume error is significantly reduced, see Fig. \ref{Fig:rt_instability position and error}(b).
%
%%%%%%%%%%%%%%%%%%%%%%%%%%%%%%%%%%%%%%%%%%%%%%
%%%%%%%%%%%%%%%%%%%%%%%%%%%%%%%%%%%%%%%%%%%%%%
%
%
%%%%%%%%%%%%%%%%%%%%%%%%%%%%%%%%%%%%%%%%%%%%%%
%%%%%%%%%%%%%%%%%%%%%%%%%%%%%%%%%%%%%%%%%%%%%%
%
\subsubsection{Bubble rising}
\label{sec4.2.3}
\begin{figure}[htbp]
\centering
\includegraphics[width=1.0\textwidth]{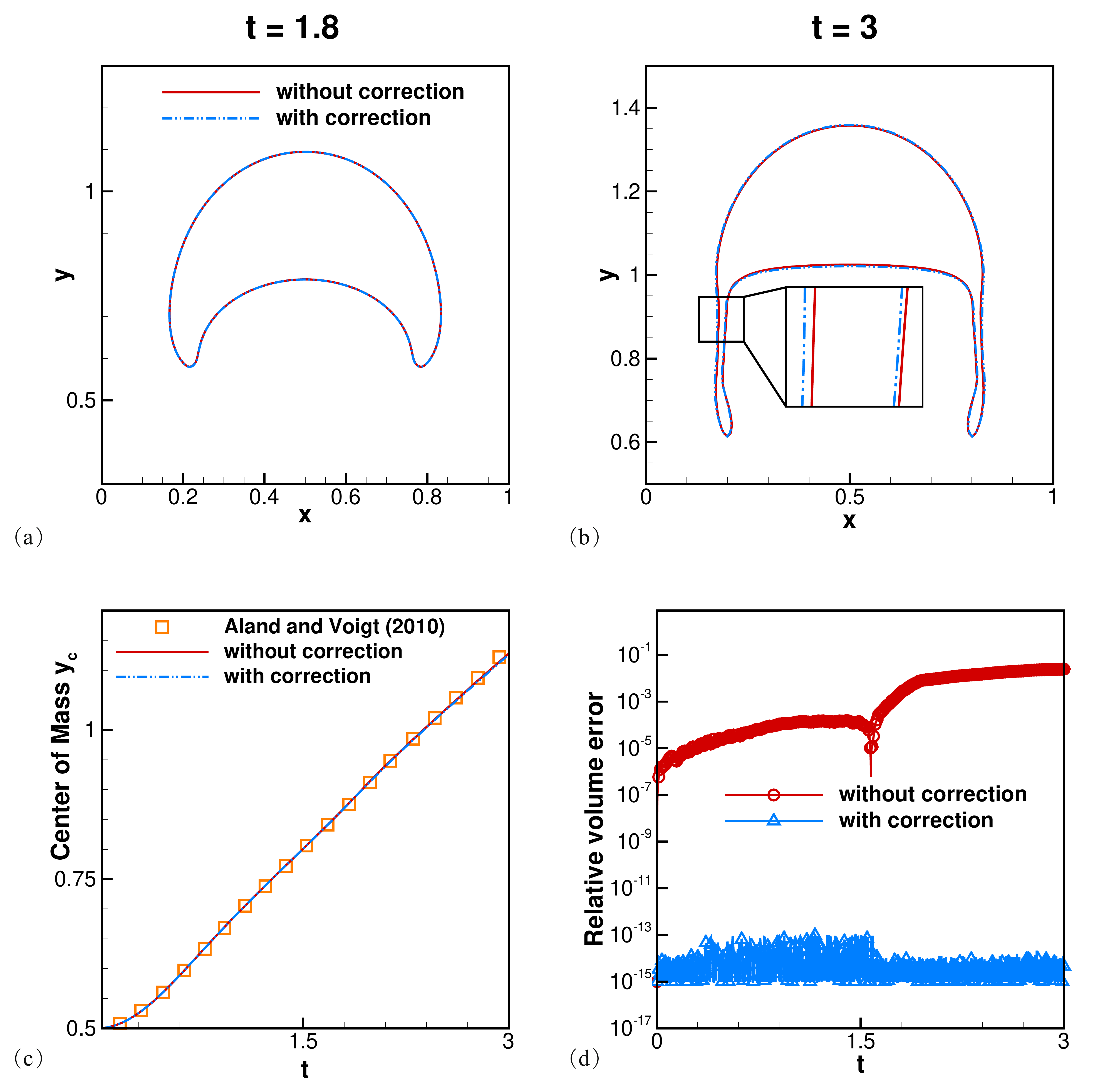}
\caption{Bubble rising (2D): (a)-(b) the evolving interfaces at different time instants, (c) the y-coordinate of the center of mass over time and (d) the relative volume error over time.}
\label{Fig:bubble rising 2D}
\end{figure}

A two-dimensional bubble rising problem is chosen here. Within a rectangular domain of $[0,1]\times[0,2]$ filled with liquid, a circular gas bubble of diameter $D = 0.5$ is initially placed at $(0.5,0.5)$ and then rises under buoyancy. The Eotvos number and the gravity Reynolds number can be used to characterize the flow filed and are defined as
\begin{equation}
\begin{aligned}
Eo &= \frac{\widetilde{\rho}_l \widetilde{g} \widetilde{D}^2}{\widetilde{\sigma}}, \\
Re_l &= \frac{\widetilde{\rho}_l \widetilde{g}^{\frac{1}{2}} \widetilde{D}^{\frac{3}{2}}}{\widetilde{\mu}_l},
\end{aligned}
\end{equation}
With $Eo = 125$ and $Re_g = 35$ \cite{yuan2018simple,aland2012benchmark}, the initial conditions are accordingly given by 
\begin{equation}
\left\{
\begin{aligned}
\rho&=1,\ u = 0, \ v = 0, \  p = 1,\ \mu = 1 \quad \text{liquid},\\
\rho&=0.001, \ u = 0, \ v = 0, \ p = 1,\ \mu = 0.01 \quad \text{gas},\\
\phi&=\sqrt{(x-0.5)^2 + (y - 0.5)^2} - 0.25\quad\text{ level\ set},\\
Re &=100,\ We = 510.2, \ Fr = 1.02 \quad\text{ dimensionless parameters},   \\ 
\end{aligned}
\right.
\end{equation}
and the termination time is $t = 3$. With a time step of $\Delta t = 3.0\times10^{-5}$, the simulations are carried out on a $128\times256$ grid using 8 cores. The CPU costs of the original method and the present method are $866.64\ \rm{s}$ and $871.86\ \rm{s}$, respectively. Since surface tension is relatively small, a pair of filaments are formed at the bottom of the bubble and become thinner as time goes on, see Fig. \ref{Fig:bubble rising 2D}(a)-(b). In Fig. \ref{Fig:bubble rising 2D}(c), the y-coordinate of the center of mass, 
\begin{equation}
y_c = \frac{\int_{\Omega}(1 - H_s(\phi)) y\ d\Omega}{V_2},
\end{equation}
%.
is plotted versus the simulation time. It is observed that the results of the present method agree well with the benchmark solution of Aland and Voigt \cite{aland2012benchmark}. The conservation of the present method is validated by the time history of the relative volume error given in Fig. \ref{Fig:bubble rising 2D}(d).  

\begin{figure}[htbp]
\centering
\includegraphics[width=1.0\textwidth]{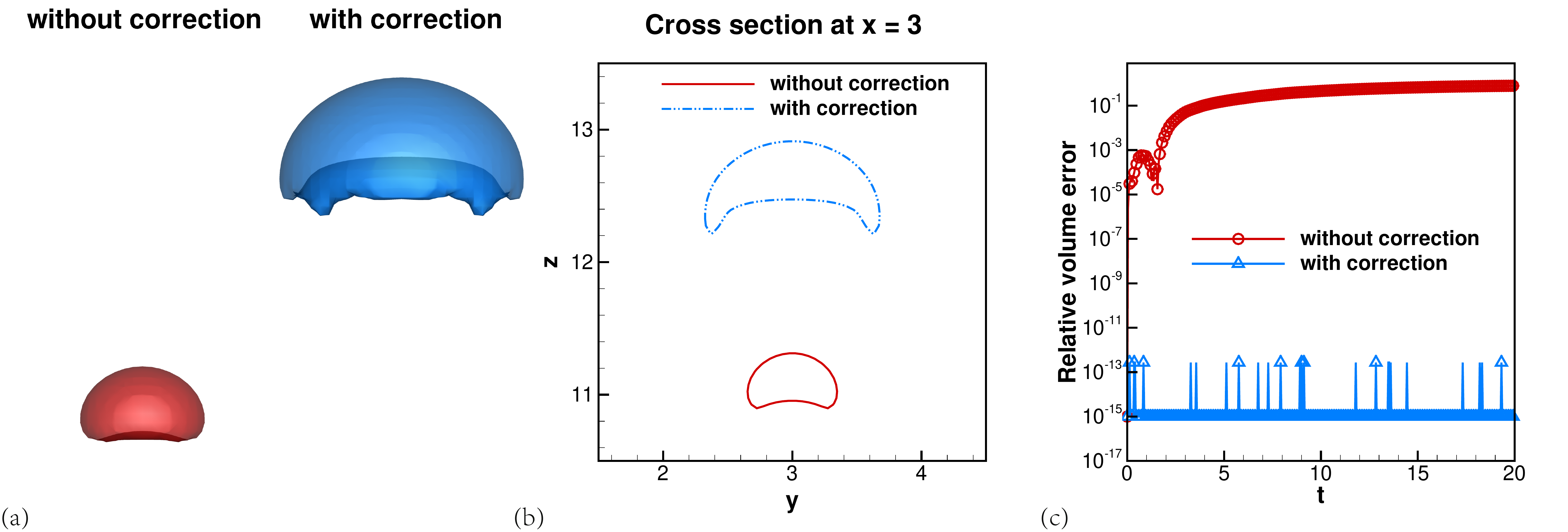}
\caption{Bubble rising (3D): (a) the 3D interface at $t = 20$, (b) the corresponding 2D cross section profile and (c) the relative volume error over time.}
\label{Fig:bubble rising 3D}
\end{figure}

Then a three-dimensional bubble rising case is computed to further validate the capability of the present method for 3D problems. By analyzing a mass of experimental data, Grace \cite{grace1973shapes} concluded that the final shapes of a single gas bubble rising in the quiescent liquid can be grouped into four categories: spherical, ellipsoidal, skirted and dimpled. The governing dimensionless parameters are the Morton number $M$, the Eotvos number $Eo$ and the terminal Reynolds number $Re_t$, which are defined as
\begin{equation}
\begin{aligned}
M &= \frac{\widetilde{g} \widetilde{\mu}_l^4 (\widetilde{\rho}_l  - \widetilde{\rho}_g)}{\widetilde{\rho}_l^2\widetilde{\sigma}^3},\\
Eo &= \frac{\widetilde{g} \widetilde{D}^2 (\widetilde{\rho}_l  - \widetilde{\rho}_g)}{\widetilde{\sigma}}, \\
Re_t &= \frac{\widetilde{\rho}_l \widetilde{U}_{\infty} \widetilde{D}}{\widetilde{\mu}_l},
\end{aligned}
\end{equation}
where $\widetilde{U}_{\infty}$ is the terminal velocity of the bubble. Note that the definition of the Eotvos number is different from that in the 2D situation. We consider the skirted case here as it is the most challenging one due to the large and rapid deformation of the bubble, see Refs. \cite{ge2018efficient,van2005numerical}. Embedded in the liquid, a spherical gas bubble of diameter $D = 1$ is initially placed at $(3,3,1.5)$ within a cubic domain of $[0,6]\times[0,6]\times[0,18]$. With $M = 0.971$ and $Eo = 97.1$, the initial conditions are given by
\begin{equation}
\left\{
\begin{aligned}
\rho&=1,\ u = 0, \ v = 0, \ w = 0, \  p = 1,\ \mu = 1 \quad \text{liquid},\\
\rho&=0.01, \ u = 0, \ v = 0, \ w = 0, \ p = 1,\ \mu = 0.01 \quad \text{gas},\\
\phi&=\sqrt{(x-3)^2 + (y - 3)^2 + (z - 1.5)^2} - 0.5\quad\text{ level\ set},\\
Re &=31.32,\ We = 98.13, \ Fr = 1 \quad\text{ dimensionless parameters}.   \\ 
\end{aligned}
\right.
\end{equation}
The termination time is set as $t = 20$ with a time step $\Delta t = 2\times10^{-3}$. With the free-slip wall boundary condition employed for the vertical walls and the no-slip wall boundary condition enforced at the horizontal walls, the computational domain is discretized on a grid involving $120\times120\times360$ cells. Obtained by using $64$ cores, the CPU time of the original method and the present method are $1762.68\ \rm{s}$ and $1776.78\ \rm{s}$, which are very close. As shown in Fig. \ref{Fig:bubble rising 3D}(a)-(b), the skirted shape is well predicted by the present method while a wrongly ellipsoidal shape is obtained by the original method owing to numerical diffusion. According to Ref. \cite{grace1973shapes}, the expected terminal Reynolds number is $20$ and the computed value of the present method is $18.3$, which is obviously better than $12.4$ predicted by the original method. The time history of the relative volume error in Fig. \ref{Fig:bubble rising 3D} also quantitatively highlights the performance of the present method. 

%
%%%%%%%%%%%%%%%%%%%%%%%%%%%%%%%%%%%%%%%%%%%%%%
%%%%%%%%%%%%%%%%%%%%%%%%%%%%%%%%%%%%%%%%%%%%%%
%

%
%%%%%%%%%%%%%%%%%%%%%%%%%%%%%%%%%%%%%%%%%%%%%%
%%%%%%%%%%%%%%%%%%%%%%%%%%%%%%%%%%%%%%%%%%%%%%
%

\subsubsection{Binary droplets collision}
\label{sec4.2.4}
\begin{figure}[htbp]
\centering
\includegraphics[width=0.8\textwidth]{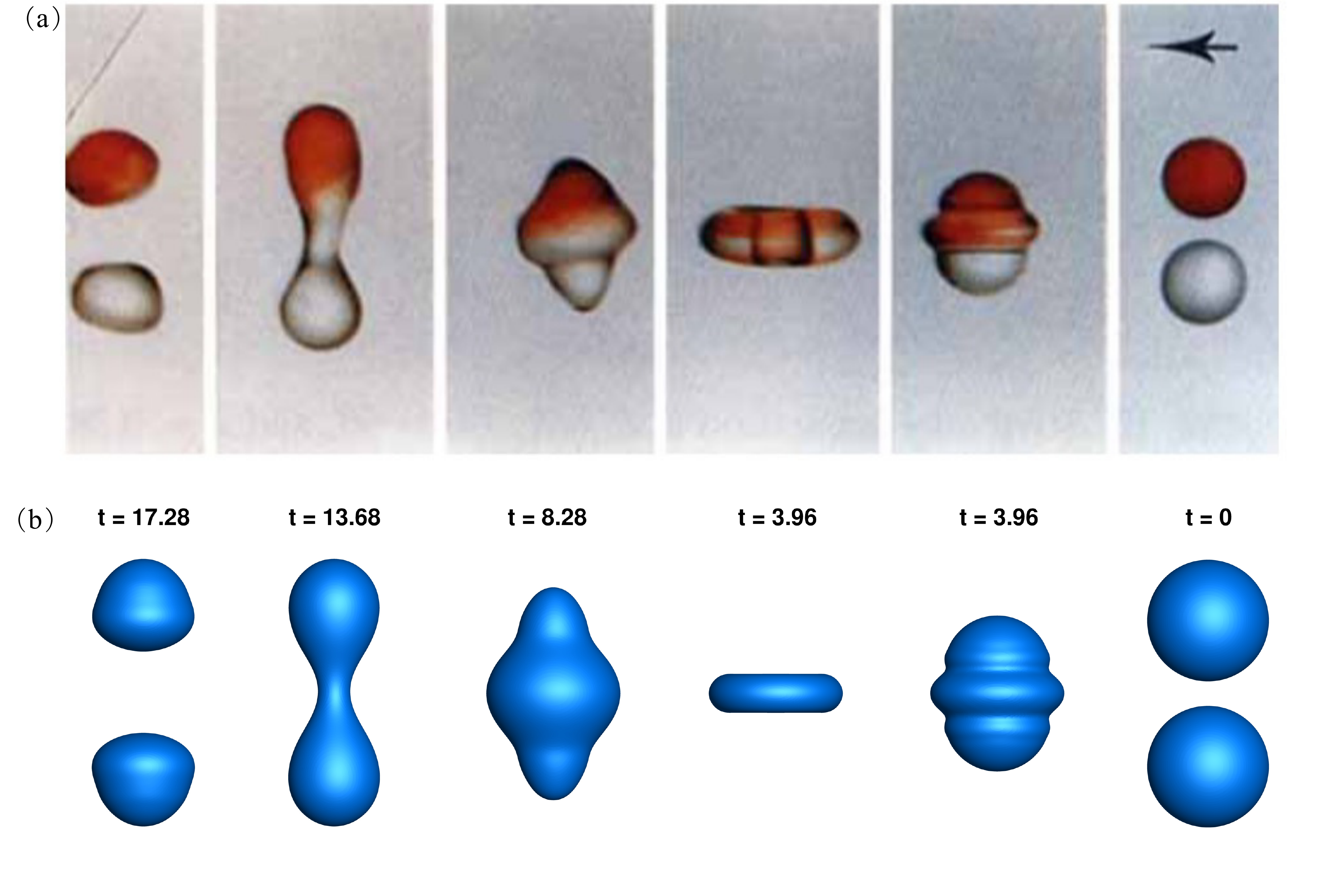}
\caption{Binary droplets collision ($We_d = 23$): (a) the experimental results taken from Ashgriz and Poo \cite{ashgriz1990coalescence} with permission of Cambridge University Press, and (b) the numerical results obtained by the present method.}
\label{Fig:binary 1}
\end{figure}

\begin{figure}[htbp]
\centering
\includegraphics[width=0.8\textwidth]{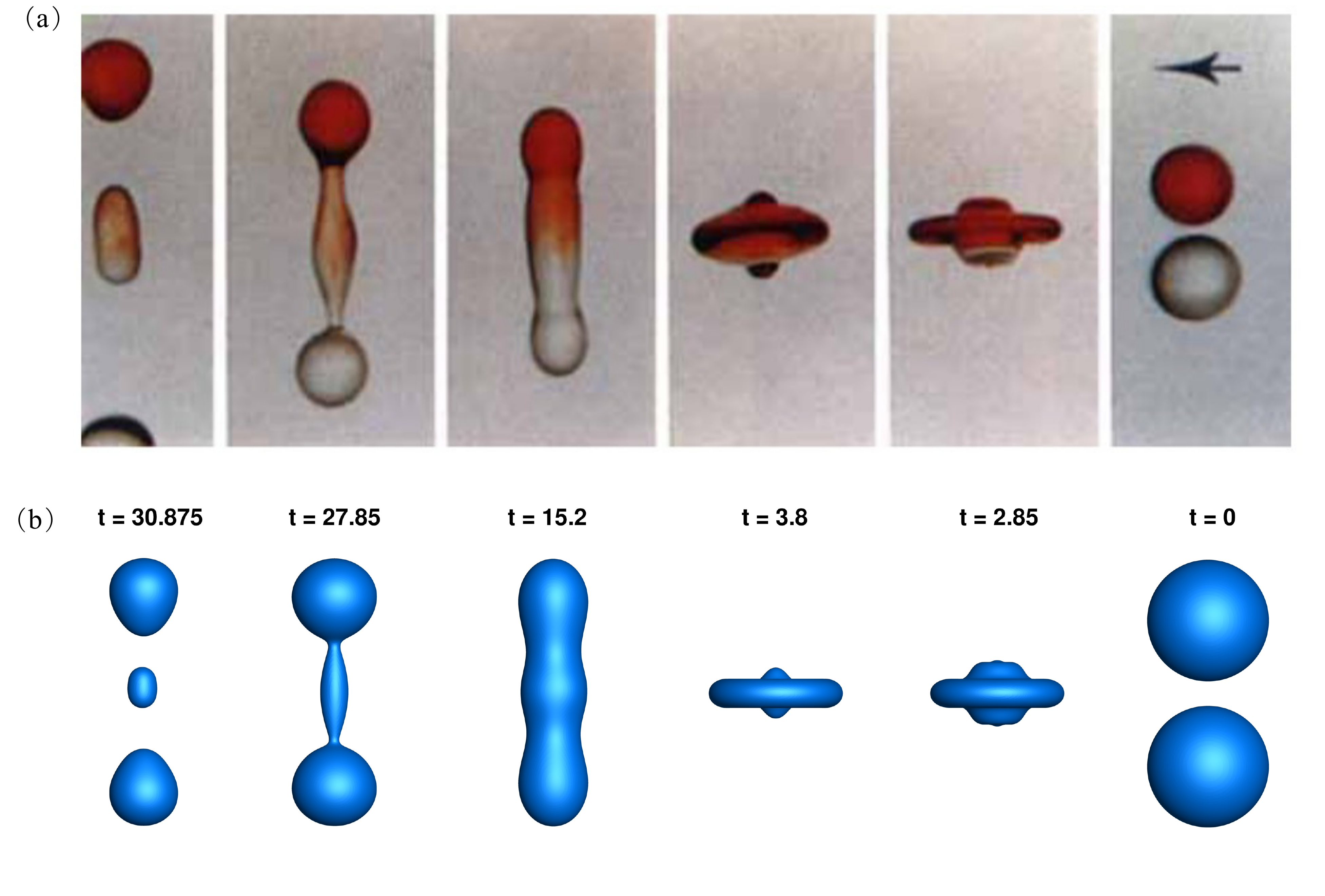}
\caption{Binary droplets collision ($We_d = 40$): (a) the experimental results taken from Ashgriz and Poo \cite{ashgriz1990coalescence} with permission of Cambridge University Press, and (b) the numerical results obtained by the present method.}
\label{Fig:binary 2}
\end{figure}

\begin{figure}[htbp]
\centering
\includegraphics[width=0.8\textwidth]{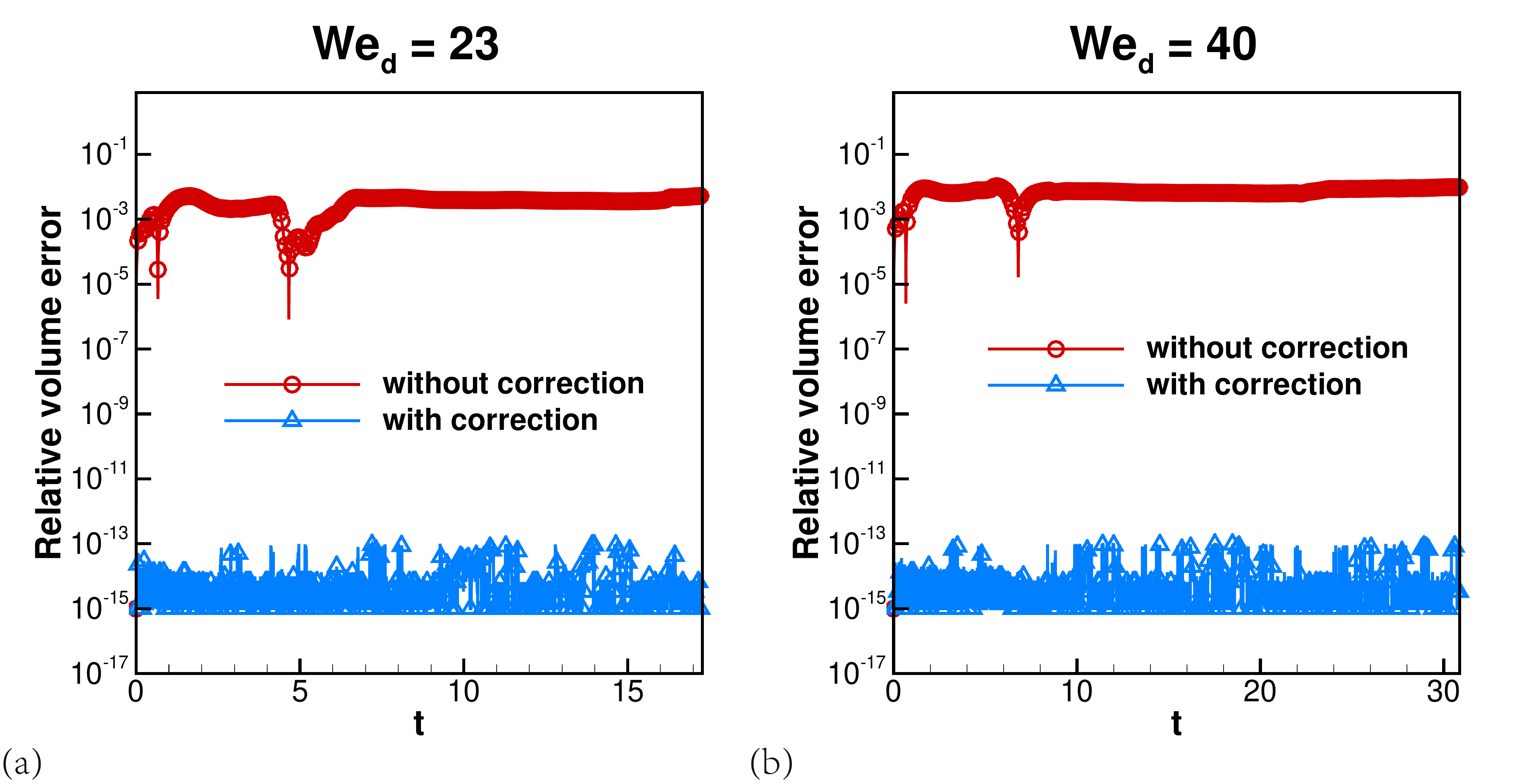}
\caption{Binary droplets collision: the relative volume error over time.}
\label{Fig:binary error}
\end{figure}

As a classic benchmark case, the three-dimensional binary droplets collision problem is considered, which has been extensively investigated \cite{luo2015massconserving,chiu2019coupled,ashgriz1990coalescence,tanguy2005application}. With no-slip wall boundary conditions, the computational domain is a cubic box of $[0,2.5D]\times [0,2.5D]\times[0,5D]$, where $D = 2.0$ is the diameter of the droplets. The dynamics of the droplets is governed by the diameter-based Weber number
\begin{equation}
We_d = \frac{\widetilde{\rho_l} \widetilde{U}_r^2 \widetilde{D}}{\widetilde{\sigma}},
\end{equation}
which is tuned by changing the relative velocity $\widetilde{U}_r$. Two situations are computed, namely $We_d = 23$ and $We_d = 40$, with the initial conditions given by
\begin{equation}
\left\{
\begin{aligned}
\rho&=63.876,\ \mu = 815.66, \  p = 1, \ u = 0, \ v = 0, \\
w&=\left\{
\begin{array}
{ll} {0.5}&\text{the\ lower\ droplet} \\ 
{-0.5}&\text{the\ upper \ droplet}
\end{array} \right. \quad \text{lquid,}\\
\rho&=1,\ \mu = 1, \ p = 1, \ u = 0, \ v = 0,\ w = 0 \quad \text{air},\\
\phi&=\sqrt{(x-2.5)^2 + (y - 2.5)^2 + (z - z_d)^2} - 0.5, \\
z_d&=\left\{
\begin{array}
{ll} {3.8}\quad &\text{the\ lower\ droplet} \\ 
{6.2}\quad  &\text{the\ upper \ droplet}
\end{array} \right. \quad\text{ level\ set,}\\
&\left\{
\begin{array}
{ll} {Re = 39.673,\ We = 0.0141\ (We_d = 23)}\\ 
{Re = 52.346,\ We = 0.0246\ (We_d = 40)}
\end{array} \right.\quad\text{dimensionless parameters.} \\
\end{aligned}
\right.
\end{equation}
The simulations are conducted on a $160\times160\times320$ grid and the time step is set to $2.5\times10^{-4}$. With $64$ cores, the simulation time of the original method and the present method are $17513.28\ \rm{s}$ and $17570.52\ \rm{s}$ for $We_d = 23$, and are $31512.96\ \rm{s}$ and $31633.92\ \rm{s}$ for $We_d = 40$. As shown in Fig. \ref{Fig:binary 1} and Fig. \ref{Fig:binary 2}, during the early stage, the evolutions of the interface for $We_d = 23$ and $We_d = 40$ are quite similar, in which a flying saucer shape is developed. Then, with the larger inertia force, a more severe bounce is observed for $We_d = 40$, resulting in an additional satellite droplet after the separation of the two droplets. The interfaces obtained by the present method match the experimental results very well. Moreover, Fig. \ref{Fig:binary error} confirms that the present method can conserve the volume even with large deformation involved.

%
%%%%%%%%%%%%%%%%%%%%%%%%%%%%%%%%%%%%%%%%%%%%%%
%%%%%%%%%%%%%%%%%%%%%%%%%%%%%%%%%%%%%%%%%%%%%%
%
%
%%%%%%%%%%%%%%%%%%%%%%%%%%%%%%%%%%%%%%%%%%%%%%
%%%%%%%%%%%%%%%%%%%%%%%%%%%%%%%%%%%%%%%%%%%%%%
%
%
%%%%%%%%%%%%%%%%%%%%%%%%%%%%%%%%%%%%%%%%%%%%%%
%%%%%%%%%%%%%%%%%%%%%%%%%%%%%%%%%%%%%%%%%%%%%%
%

\section{Concluding remarks}
\label{sec5}
In this paper, a novel mass correction procedure has been developed to overcome the uphysical mass gain/loss in the level-set method. Inspired by the scale-separation algorithm of Han et al. \cite{han2015scale}, we introduce a small perturbation to the level-set filed to eliminate the conservation error, which can be solved by using the Newton method. With the signed distance property of the level-set function preserved exactly, the correction procedure can be implemented after the reinitialization step. Thus, unlike in previous researches \cite{yuan2018simple, luo2015massconserving, ge2018efficient}, no additional conservation error will be reintroduced. In addition, the present method is easy to implement in both 2D and 3D simulations since only algebraic operations are needed. To validate the present method, a number of benchmark cases involving large deformations are computed. With the present method, the mass is well-conserved and more details in the flow field are captured. Quantitative study indicates that the conservation error is reduced to the order of machine accuracy by employing the present method and the extra computational cost is negligible. In future work, we plan to add a phase change model in the present method to simulate the icing of droplets.

\section{Acknowledgements}

%\appendix

%\section{Section in Appendix}
%\label{appendix-sec1}

%Sample text. Sample text. Sample text. Sample text. Sample text. Sample text. 
%Sample text. Sample text. Sample text. Sample text. Sample text. Sample text. 
%Sample text. 

%% References


%\bibliographystyle{elsarticle-num}
% \bibliographystyle{elsarticle-harv}
% \bibliographystyle{elsarticle-num-names}
% \bibliographystyle{model1a-num-names}
% \bibliographystyle{model1b-num-names}
% \bibliographystyle{model1c-num-names}
% \bibliographystyle{model2-names}
% \bibliographystyle{model3a-num-names}
% \bibliographystyle{model3-num-names}
% \bibliographystyle{model4-names}
% \bibliographystyle{model5-names}
% \bibliographystyle{model6-num-names}

\bibliographystyle{unsrt}
\bibliography{ref}
\end{document}